\documentclass[10pt,a4paper]{article}

\usepackage[utf8]{inputenc}
\usepackage[english]{babel}
\usepackage{amssymb,amsmath}
\usepackage{euscript}

\begin{document}
\large

\newpage
\begin{center}
\LARGE{\bf CP-Symmetry in Scattering of Neutrinos from Nuclei}
\end{center}
\vspace{0.1mm}
\begin{center}
{\bf R.B. Begzhanov and R.S. Sharafiddinov}
\end{center}
\vspace{0.1mm}
\begin{center}
{\bf Institute of Nuclear Physics, Uzbekistan Academy of Sciences,
\\Ulugbek, Tashkent 100214, Uzbekistan}
\end{center}
\vspace{0.1mm}

\begin{center}
{\bf Abstract}
\end{center}

The elastic scattering of longitudinal and transversal neutrinos on a spinless
nucleus have been discussed taking into account the charge, magnetic, anapole
and electric dipole moments of fermions and their weak neutral currents.
Compound structure of the neutrino interaction cross section with nuclei have
been defined. Invariance of the considered process concerning the C - and
P-operations have been investigated in the polarization type dependence.

\vspace{0.8cm}
\noindent
{\bf 1. Introduction}
\vspace{0.4cm}

It has been established that the behavior of massive neutrinos plays an
important part in understanding the physical nature of elementary particles.
One of the modes of doing this is to study the possible neutrino-nucleus
interaction [1,2].

The neutrino interaction with field of emission may be expressed in the
form [3,4] of electromagnetic current
$$j_{\mu}^{em}=\overline{u}(p',s')[\gamma_{\mu}F_{1\nu}(q^{2})-
i\sigma_{\mu\lambda}q_{\lambda}F_{2\nu}(q^{2})+$$
\begin{equation}
+\gamma_{5}\gamma_{\mu}G_{1\nu}(q^{2})-
i\gamma_{5}\sigma_{\mu\lambda}q_{\lambda}G_{2\nu}(q^{2})]u(p,s),
\label{1}
\end{equation}
where $\sigma_{\mu\lambda}=[\gamma_{\mu},\gamma_{\lambda}]/2,$ $q=p-p'$
is the momentum transfer, $p(s)$ and $p'(s')$ imply the four-momentum
(helicities) of initial and final neutrinos, $F_{i\nu}(q^{2})$ and
$G_{i\nu}(q^{2})$ are the interaction vector and axial-vector parts
respectively. The functions $F_{1\nu}(0),$ $F_{2\nu}(0)$ and $G_{2\nu}(0)$
give the static estimates of the neutrino charge [5], magnetic [6]
and electric dipole [7] moments, on which there exist experimental and
cosmological bounds [8]. Insofar as $G_{1\nu}(0)$ is concerned, it
defines the size of the anapole moment [9], but its value has not
yet been measured in the laboratory [10].

It is known that $F_{i\nu}(q^{2})$ are invariant with respect to C - and
P-operations because the interaction of $F_{i\nu}(q^{2})$ with field of
emission must be CP-symmetrical. The term $G_{1\nu}(q^{2})$ is CP-even
but P-odd [9]. In contrast to this, the term $G_{2\nu}(q^{2})$ must
be C-invariant but CP-antisymmetrical [11]. Therefore, the form
factors $G_{i\nu}(q^{2})$ may be different from zero only
in the case where P-symmetry is absent.

The violation of P-parity leads to the appearance of right-left asymmetry,
for example, at the polarized neutrinos scattering on nuclei. In many works
[2,12] the spin phenomena was studied with longitudinal neutrinos. Such
an investigation is important not only for elucidation of compound structure
of the interaction between leptons and hadrons but also for observation and
refinement of the most diverse symmetries of elementary particles. However,
the massive neutrino must have the longitudinal as well as the transversal
polarization. The account of the latter gives the possibility to directly
look at the nature of the discussed processes.

In the present work, we investigate the phenomena of symmetricality
in the massive neutrinos interactions with an electroweak field of emission.
Section 2 is dedicated to the elastic scattering of longitudinal polarized
neutrinos on the nucleus electric $(Z)$ and weak $(Z_{W})$ charges
\begin{equation}
\nu(\overline{\nu})+A(Z, Z_{W})\stackrel{\gamma,Z^{0}}
{\rightarrow}\nu'(\overline{\nu'})+ A(Z, Z_{W}),
\label{2}
\end{equation}
going at the expense of neutral and electromagnetic currents. In Sec. 3
the studied processes have been reanalysed for the transversal case of the
neutrino polarization. In Sec. 4 we make some concluding remarks.

\vspace{0.8cm}
\noindent
{\bf 2. Longitudinal Polarized Neutrinos Scattering on a Nucleus}
\vspace{0.4cm}

In the framework of the standard theory of electroweak interaction [13],
the Hamiltonian of the neutrino interaction with field of a nucleus
has the form
$$H=\frac{4\pi\alpha}{q^{2}}\overline{u}(p',s')[\gamma_{\mu}F_{1\nu}(q^{2})-
i\sigma_{\mu\lambda}q_{\lambda}F_{2\nu}(q^{2})+$$
$$+\gamma_{5}\gamma_{\mu}G_{1\nu}(q^{2})-
i\gamma_{5}\sigma_{\mu\lambda}q_{\lambda}G_{2\nu}(q^{2})]
u(p,s)J_{\mu}^{\gamma}(q)+$$
\begin{equation}
+\frac{G_{F}}{\sqrt{2}}\overline{u}(p',s')\gamma_{\mu}(g_{V_{\nu}}+
\gamma_{5}g_{A_{\nu}})u(p,s)J_{\mu}^{Z^{0}}(q).
\label{3}
\end{equation}
Here $J_{\mu}^{x}(q)$ are the nucleus electromagnetic $(x=\gamma)$ and
weak neutral $(x=Z^{0})$ currents [14], $g_{V_{\nu}}$ and $g_{A_{\nu}}$
are the corresponding constants of the neutrino interaction vector $(V)$ and
axial $(A)$ parts.

In the case of the neutrino longitudinl polarization and of a zero-spin
nucleus, the cross-section of the process (\ref{2}) on the basis of (\ref{3})
can be presented after the summing of $s'$ as follows:
\begin{equation}
d\sigma_{ew}(\theta,s)=
d\sigma_{em}(\theta,s)+d\sigma_{int}(\theta,s)+d\sigma_{we}(\theta,s),
\label{4}
\end{equation}
where to purely electromagnetic interaction answers the expression
$$\frac{d\sigma_{em}(\theta,s)}{d\Omega}=
\sigma^{\nu}_{o}(1-\eta^{2}_{\nu})^{-1}\{[F_{1\nu}+
2\lambda_{c}s\sqrt{1-\eta_{\nu}^{2}}G_{1\nu}]F_{1\nu}+$$
$$+\eta^{2}_{\nu}[F_{1\nu}^{2}+
4m_{\nu}^{2}(1-\eta^{-2}_{\nu})^{2}F_{2\nu}^{2}]tg^{2}\frac{\theta}{2}-
8sE_{\nu}^{2}(1-\eta_{\nu}^{2})^{3/2}F_{2\nu}G_{2\nu}
tg^{2}\frac{\theta}{2}+$$
\begin{equation}
+(1-\eta^{2}_{\nu})[G_{1\nu}^{2}+
4E_{\nu}^{2}G_{2\nu}^{2}tg^{2}\frac{\theta}{2}]\}F_{E}^{2}(q^{2}).
\label{5}
\end{equation}

The contribution explained by the interference of electroweak interaction
is written in the form
$$\frac{d\sigma_{int}(\theta,s)}{d\Omega}=
\rho\sigma^{\nu}_{o}(1-\eta^{2}_{\nu})^{-1}g_{V_{\nu}}\{[1-$$
\begin{equation}
-\lambda_{c}s\frac{g_{A_{\nu}}}{g_{V_{\nu}}}\sqrt{1-
\eta^{2}_{\nu}}][F_{1\nu}+
\lambda_{c}s\sqrt{1-\eta_{\nu}^{2}}G_{1\nu}]+
\eta^{2}_{\nu}F_{1\nu}tg^{2}\frac{\theta}{2}\}F_{EV}(q^{2}).
\label{6}
\end{equation}

In the same way one can present the cross-section of purely weak
interaction with neutral currents
$$\frac{d\sigma_{we}(\theta,s)}{d\Omega}=
\frac{E_{\nu}^{2}G_{F}^{2}}{8\pi^{2}}
\{g_{V_{\nu}}^{2}(1+\eta_{\nu}^{2}tg^{2}\frac{\theta}{2})+
g_{A_{\nu}}^{2}(1-\eta_{\nu}^{2})-$$
\begin{equation}
-2\lambda_{c}sg_{V_{\nu}}g_{A_{\nu}}
\sqrt{1-\eta_{\nu}^{2}}\}F_{W}^{2}(q^{2})cos^{2}\frac{\theta}{2}.
\label{7}
\end{equation}
Here we have also used the size
$$\sigma_{o}^{\nu}=
\frac{\alpha^{2}cos^{2}\frac{\theta}{2}}{4E_{\nu}^{2}(1-\eta^{2}_{\nu})
sin^{4}\frac{\theta}{2}}, \, \, \, \,
\eta_{\nu}=\frac{m_{\nu}}{E_{\nu}}, \, \, \, \,
\rho=\frac{G_{F}q^{2}}{2\pi\sqrt{2}\alpha},$$
$$F_{E}(q^{2})=ZF_{c}(q^{2}), \, \, \, \,
F_{EV}(q^{2})=ZZ_{W}F_{c}^{2}(q^{2}), \, \, \, \, F_{W}(q^{2})=
Z_{W}F_{c}(q^{2}),$$
$$Z_{W}=\frac{1}{2}\{\beta_{V}^{(0)}(Z+N)+\beta_{V}^{(1)}
(Z-N)\}, \, \, \, \, A=Z+N, \, \, \, \, M_{T}=\frac{1}{2}(Z-N),$$
where $\theta$ is the scattering angle, $E_{\nu}$ and $m_{\nu}$ are the
neutrino mass and energy, $F_{c}(q^{2})$ is the charge ($F_{c}(0)=1$) form
factor of a nucleus with isospin T and its projection $M_{T},\beta_{V}^{(0)}$
and $\beta_{V}^{(1)}$ are constants of isoscalar and isovector components
of vector neutral hadronic current.

The presence of the multiplier $s$ in Eqs. (\ref{5})-(\ref{7}) implies their
antisymmetricality concerning the substitution of the left-handed $(s=-1)$
particle with the right-handed $(s=+1)$ and vice versa. We see in addition
that Eqs. (\ref{5})-(\ref{7}) for the neutrino $(\lambda_{c}=+1)$ and the
antineutrino $(\lambda_{c}=-1)$ are different.

Taking into account Eqs. (\ref{5})-(\ref{7}), the size of charge asymmetry
\begin{equation}
A_{ch}^{ew}=A_{ch}^{em}+A_{ch}^{int}+A_{ch}^{we}=
\frac{d\sigma^{\nu}_{ew}-d\sigma^{\overline{\nu}}_{ew}}
{d\sigma^{\nu}_{ew}+d\sigma^{\overline{\nu}}_{ew}}
\label{8}
\end{equation}
is defined by the corresponding contributions
$$A_{ch}^{em}(\theta)=
2s\sqrt{1-\eta^{2}_{\nu}}F_{1\nu}G_{1\nu}
\{(1+\eta^{2}_{\nu}tg^{2}\frac{\theta}{2})F_{1\nu}^{2}+$$
\begin{equation}
+(1-\eta^{2}_{\nu})G_{1\nu}^{2}+
4E_{\nu}^{2}[s\sqrt{1-\eta_{\nu}^{2}}G_{2\nu}-
(1-\eta^{2}_{\nu})F_{2\nu}]^{2}tg^{2}\frac{\theta}{2}\}^{-1},
\label{9}
\end{equation}
$$A_{ch}^{int}(\theta)=
s\sqrt{1-\eta_{\nu}^{2}}[G_{1\nu}-\frac{g_{A_{\nu}}}{g_{V_{\nu}}}F_{1\nu}]
\{(1+\eta^{2}_{\nu}tg^{2}\frac{\theta}{2})F_{1\nu}-$$
\begin{equation}
-\frac{g_{A_{\nu}}}{g_{V_{\nu}}}(1-\eta^{2}_{\nu})G_{1\nu}\}^{-1},
\label{10}
\end{equation}
\begin{equation}
A_{ch}^{we}(\theta)=-2s\frac{g_{A_{\nu}}}{g_{V_{\nu}}}
\sqrt{1-\eta_{\nu}^{2}}\{(1+\eta_{\nu}^{2}tg^{2}\frac{\theta}{2})+
\frac{g_{A_{\nu}}^{2}}{g_{V_{\nu}}^{2}}(1-\eta_{\nu}^{2})\}^{-1}.
\label{11}
\end{equation}

These formulas show clearly that C-invariance of the considered process
can be violated only in the case when the mirror symmetry is absent.
Indeed, taking $s=0,$ we find
\begin{equation}
A_{ch}^{em}(\theta)=0, \, \, \, \, A_{ch}^{int}
(\theta)=0, \, \, \, \, A_{ch}^{we}(\theta)=0,
\label{12}
\end{equation}
which are true at the conservation of P-parity.

Many authors state that one must use the electromagnetic current (\ref{1})
in the form [15] in which an $i$ is absent. If we start with such a
procedure, assuming that the interaction magnetic and electric dipole terms
must not be Hermitian even with $q^{2}<0,$ we would establish the other
expressions for the processes cross-sections instead of (\ref{5}) and
(\ref{6}). They lead to the implication [16] that C-invariance of
elastic scattering is basically violated at the expense of the neutrino
nonzero rest mass. One can also make a conclusion that this influence
does not relate to the behavior of P-symmetry.

Taking into account that nonconservation of P-parity at the neutrino
interaction conveniently characterize by the right-left asymmetry, we have
\begin{equation}
A_{RL}^{ew}=A_{RL}^{em}+A_{RL}^{int}+A_{RL}^{we}=
\frac{d\sigma_{ew}^{R}-d\sigma_{ew}^{L}}
{d\sigma_{ew}^{R}+d\sigma_{ew}^{L}},
\label{13}
\end{equation}
from Eqs. (\ref{5})-(\ref{7}), we get
$$A_{RL}^{em}(\theta)=
2\sqrt{1-\eta_{\nu}^{2}}[\lambda_{c}F_{1\nu}G_{1\nu}-$$
$$-4E_{\nu}^{2}(1-\eta^{2}_{\nu})F_{2\nu}G_{2\nu}
tg^{2}\frac{\theta}{2}]\{(1+\eta^{2}_{\nu}tg^{2}\frac{\theta}{2})
F_{1\nu}^{2}+$$
\begin{equation}
+(1-\eta_{\nu}^{2})[G_{1\nu}^{2}ctg^{2}\frac{\theta}{2}+
4E_{\nu}^{2}(G_{2\nu}^{2}+(1-\eta^{2}_{\nu})F_{2\nu}^{2})]
tg^{2}\frac{\theta}{2}\}^{-1},
\label{14}
\end{equation}
$$A_{RL}^{int}(\theta)=
\lambda_{c}\sqrt{1-\eta_{\nu}^{2}}[G_{1\nu}-
\frac{g_{A_{\nu}}}{g_{V_{\nu}}}F_{1\nu}]\{(1+
\eta^{2}_{\nu}tg^{2}\frac{\theta}{2})F_{1\nu}-$$
\begin{equation}
-\frac{g_{A_{\nu}}}{g_{V_{\nu}}}(1-\eta^{2}_{\nu})G_{1\nu}\}^{-1},
\label{15}
\end{equation}
\begin{equation}
A_{RL}^{we}(\theta)=-2\lambda_{c}\frac{g_{A_{\nu}}}{g_{V_{\nu}}}
\sqrt{1-\eta_{\nu}^{2}}\{(1+\eta_{\nu}^{2}tg^{2}\frac{\theta}{2})+
\frac{g_{A_{\nu}}^{2}}{g_{V_{\nu}}^{2}}(1-\eta_{\nu}^{2})\}^{-1}.
\label{16}
\end{equation}

The availability of the multiplier $\lambda_{c}$ in these formulas
implies the influence of the interaction C-antisymmetrical structure
on the conservation of P-symmetry. Indeed, the average cross-sections,
Eqs. (\ref{5})-(\ref{7}), over the two values of $\lambda_{c}$ would leads
us to the equalities
$$A_{RL}^{em}(\theta)=
-8E_{\nu}^{2}(1-\eta^{2}_{\nu})^{3/2}F_{2\nu}G_{2\nu}
\{(1+\eta^{2}_{\nu}tg^{2}\frac{\theta}{2})F_{1\nu}^{2}+$$
\begin{equation}
+(1-\eta_{\nu}^{2})[G_{1\nu}^{2}+
4E_{\nu}^{2}(G_{2\nu}^{2}+(1-\eta^{2}_{\nu})F_{2\nu}^{2})
tg^{2}\frac{\theta}{2}]\}^{-1}
tg^{2}\frac{\theta}{2},
\label{17}
\end{equation}
\begin{equation}
A_{RL}^{int}(\theta)=0, \, \, \, \, A_{RL}^{we}(\theta)=0,
\label{18}
\end{equation}
which take place at C-invariance.

Thus, it follows that regardless of the behavior of charge symmetry, the
right-left asymmetry of the process (\ref{2}) can be explained by the
interference of the interaction axial-vector terms with its vector terms,
if neutrinos do not possess any new properties.

\vspace{0.8cm}
\noindent
{\bf 3. Interaction of Transversal Polarized Neutrinos 
\\with an Electroweak Field of a Nucleus}
\vspace{0.4cm}

Starting from (\ref{3}) and assuming that the neutrinos are strictly
transversal, for the elastic scattering cross-section we find an explicit
expression which can be reduced after the summing of $s'$ to the form
\begin{equation}
d\sigma_{ew}(\theta,\varphi,s)=
d\sigma_{em}(\theta,\varphi,s)+d\sigma_{int}(\theta,\varphi,s)+
d\sigma_{we}(\theta,\varphi,s).
\label{19}
\end{equation}

As well as in (\ref{4}), each term here corresponds to the most diverse
process and has the different structure:
$$\frac{d\sigma_{em}(\theta,\varphi,s)}{d\Omega}=
\sigma^{\nu}_{o}(1-\eta^{2}_{\nu})^{-1}\{F_{1\nu}^{2}+
\eta^{2}_{\nu}[F_{1\nu}^{2}+4m_{\nu}^{2}(1-\eta^{-2}_{\nu})^{2}F_{2\nu}^{2}]
tg^{2}\frac{\theta}{2}+$$
$$+2\lambda_{c}s\eta_{\nu}\sqrt{1-\eta_{\nu}^{2}}F_{1\nu}
G_{1\nu}tg\frac{\theta}{2}cos^{2}\varphi+$$
\begin{equation}
+(1-\eta^{2}_{\nu})[G_{1\nu}^{2}+
4E_{\nu}^{2}G_{2\nu}^{2}tg^{2}\frac{\theta}{2}]\}F_{E}^{2}(q^{2}),
\label{20}
\end{equation}
$$\frac{d\sigma_{int}(\theta,\varphi,s)}{d\Omega}=
\rho\sigma^{\nu}_{o}(1-\eta^{2}_{\nu})^{-1}g_{V_{\nu}}\{F_{1\nu}+
\eta_{\nu}^{2}[1+$$
$$+\lambda_{c}s\frac{g_{A_{\nu}}}{g_{V_{\nu}}}\eta_{\nu}^{-1}
\sqrt{1-\eta^{2}_{\nu}}ctg\frac{\theta}{2}cos^{2}\varphi]F_{1\nu}
tg^{2}\frac{\theta}{2}-$$
\begin{equation}
-\lambda_{c}s\eta_{\nu}\sqrt{1-\eta^{2}_{\nu}}
[tg\frac{\theta}{2}cos^{2}\varphi+
\lambda_{c}s\frac{g_{A_{\nu}}}{g_{V_{\nu}}}\eta_{\nu}^{-1}
\sqrt{1-\eta^{2}_{\nu}}]G_{1\nu}\}F_{EV}(q^{2}),
\label{21}
\end{equation}
$$\frac{d\sigma_{we}(\theta,\varphi,s)}{d\Omega}=
\frac{E_{\nu}^{2}G_{F}^{2}}{8\pi^{2}}
\{g_{V_{\nu}}^{2}(1+\eta_{\nu}^{2}tg^{2}\frac{\theta}{2})+
g_{A_{\nu}}^{2}(1-\eta_{\nu}^{2})-$$
\begin{equation}
-2\lambda_{c}sg_{V_{\nu}}g_{A_{\nu}}\eta_{\nu}
\sqrt{1-\eta_{\nu}^{2}}tg\frac{\theta}{2}cos^{2}\varphi\}
F_{W}^{2}(q^{2})cos^{2}\frac{\theta}{2},
\label{22}
\end{equation}
where $\varphi$ is the azimuthal angle.

Using (\ref{20})-(\ref{22}) and taking (\ref{8}), for the C-odd asymmetry
in the case of the neutrino transversal polarization we get
$$A_{ch}^{em}(\theta,\varphi)=
2s\eta_{\nu}\sqrt{1-\eta_{\nu}^{2}}F_{1\nu}G_{1\nu}
\{(1+\eta^{2}_{\nu}
tg^{2}\frac{\theta}{2})F_{1\nu}^{2}+$$
\begin{equation}
+(1-\eta^{2}_{\nu})[G_{1\nu}^{2}+
4E_{\nu}^{2}(G_{2\nu}^{2}+
(1-\eta^{2}_{\nu})F_{2\nu}^{2})tg^{2}\frac{\theta}{2}]\}^{-1}
tg\frac{\theta}{2}cos^{2}\varphi,
\label{23}
\end{equation}
$$A_{ch}^{int}(\theta,\varphi)=
-s\eta_{\nu}\sqrt{1-\eta_{\nu}^{2}}
[G_{1\nu}-\frac{g_{A_{\nu}}}{g_{V_{\nu}}}F_{1\nu}]
\{(1+\eta^{2}_{\nu}tg^{2}\frac{\theta}{2})F_{1\nu}-$$
\begin{equation}
-\frac{g_{A_{\nu}}}{g_{V_{\nu}}}(1-\eta^{2}_{\nu})G_{1\nu}\}^{-1}
tg\frac{\theta}{2}cos^{2}\varphi,
\label{24}
\end{equation}
$$A_{ch}^{we}(\theta,\varphi)=-2s\eta_{\nu}\frac{g_{A_{\nu}}}{g_{V_{\nu}}}
\sqrt{1-\eta_{\nu}^{2}}\{(1+\eta_{\nu}^{2}tg^{2}\frac{\theta}{2})+$$
\begin{equation}
+\frac{g_{A_{\nu}}^{2}}{g_{V_{\nu}}^{2}}(1-\eta_{\nu}^{2})\}^{-1}
tg\frac{\theta}{2}cos^{2}\varphi.
\label{25}
\end{equation}

The solutions (\ref{23})-(\ref{25}) at $s=0$ coincide with the corresponding
size from (\ref{12}) and that, consequently, the behavior of C-invariance
in the P-symmetrical interactions does not depend on the type of polarization.

In the same way one can see that the P-odd characteristics of elastic
scattering, according to (\ref{13}), (\ref{20})-(\ref{22}), has the form
$$A_{RL}^{em}(\theta,\varphi)=
2\lambda_{c}\eta_{\nu}\sqrt{1-\eta_{\nu}^{2}}F_{1\nu}G_{1\nu}
\{(1+\eta^{2}_{\nu}tg^{2}\frac{\theta}{2})F_{1\nu}^{2}+$$
\begin{equation}
+(1-\eta_{\nu}^{2})[G_{1\nu}^{2}ctg^{2}\frac{\theta}{2}+
4E_{\nu}^{2}(G_{2\nu}^{2}+(1-\eta^{2}_{\nu})F_{2\nu}^{2})]
tg^{2}\frac{\theta}{2}\}^{-1}tg\frac{\theta}{2}cos^{2}\varphi,
\label{26}
\end{equation}
$$A_{RL}^{int}(\theta,\varphi)=
-\lambda_{c}\eta_{\nu}\sqrt{1-\eta_{\nu}^{2}}[G_{1\nu}-
\frac{g_{A_{\nu}}}{g_{V_{\nu}}}F_{1\nu}]
\{(1+\eta^{2}_{\nu}tg^{2}\frac{\theta}{2})F_{1\nu}-$$
\begin{equation}
-\frac{g_{A_{\nu}}}{g_{V_{\nu}}}(1-\eta^{2}_{\nu})G_{1\nu}\}^{-1}
tg\frac{\theta}{2}cos^{2}\varphi,
\label{27}
\end{equation}
$$A_{RL}^{we}(\theta,\varphi)=-2\lambda_{c}\eta_{\nu}\frac{g_{A_{\nu}}}
{g_{V_{\nu}}}\sqrt{1-\eta_{\nu}^{2}}\{(1+\eta_{\nu}^{2}
tg^{2}\frac{\theta}{2})+$$
\begin{equation}
+\frac{g_{A_{\nu}}^{2}}{g_{V_{\nu}}^{2}}(1-\eta_{\nu}^{2})\}^{-1}
tg\frac{\theta}{2}cos^{2}\varphi.
\label{28}
\end{equation}

However, due to C-parity, it follows from (\ref{26})-(\ref{28}) that
\begin{equation}
A_{RL}^{em}(\theta,\varphi)=0, \, \, \, \, A_{RL}^{int}
(\theta,\varphi)=0, \, \, \, \, A_{RL}^{we}(\theta,\varphi)=0.
\label{29}
\end{equation}

Comparing (\ref{17}) and (\ref{18}) with (\ref{29}), it is easy to observe
the differences which may serve as an indication to the type of polarization
dependence of C-invariant processes right-left asymmetry.

\vspace{0.8cm}
\noindent
{\bf 4. Conclusion}
\vspace{0.4cm}

We have established an explicit form of the differential cross sections
describing the elastic electroweak scattering of longitudinal and transversal
polarized neutrinos (antineutrinos) on spinless nuclei as a consequence of
the availability of rest mass, charge, magnetic, anapole and electric dipole
moments of elementary particles and their weak neutral currents. With the use
of these formulas a proof has been obtained regardless of the nature of C,
nonconservation of P can be explained by the interference of the interaction
vector and axial-vector parts.

One of the new features of our results is the connection between the P-odd
phenomena and possible polarization types. Unlike the behavior of C-parity
in the P-symmetrical scattering, coefficients of right-left asymmetries
$A_{RL}^{ew}(\theta)$ and $A_{RL}^{ew}(\theta,\varphi)$ in the C-invariant
processes with longitudinal and transversal neutrinos are different.

Furthermore, if neutrinos are of high energies $(E_{\nu}\gg m_{\nu})$ then
$A_{RL}^{ew}(\theta,\varphi)=0,$ and the size of $A_{RL}^{ew}(\theta)$
is reduced to the form
\begin{equation}
A_{RL}^{ew}(\theta)=-\frac{2F_{2\nu}G_{2\nu}}{F_{2\nu}^{2}+G_{2\nu}^{2}}.
\label{30}
\end{equation}

It is expected that measurement of right-left asymmetry $A_{RL}^{ew}(\theta)$
for any two values of large energies will testify in favor of the equality
of the neutrino magnetic and electric dipole moments.

\vspace{0.8cm}
\noindent
{\bf References}
\begin{enumerate}
\item
P.H. Frampton and P. Vogel, {\it Phys. Rep.} {\bf 82},
339 (1982); F. Boehm and P. Vogel, {\it Ann. Rev. Nucl. Part. Sci.}
{\bf 34}, 125 (1984); P. Vogel and J. Engel, {\it Phys. Rev.}
{\bf D39}, 3378 (1989).
\item 
B.K. Kerimov, T.R. Aruri and M.Ya. Safin, {\it Izv. Acad.
Nauk SSSR. Ser. Fiz.} {\bf 37}, 1768 (1973); B.K. Kerimov and M.Ya. Safin,
{\it Izv. Russ. Acad. Nauk Ser. Fiz.} {\bf 61}, 657 (1997).
\item 
M.A.B. Beg, W.J. Marciano and M. Ruderman, {\it Phys.
Rev.} {\bf D17}, 1395 (1978).
\item
W. Bernreuther and M. Suzuki, {\it Rev. Mod. Phys.}
{\bf 63}, 313 (1991).
\item 
R.S. Sharafiddinov, {\it Dokl. Akad. Nauk Ruz. Ser. Math.
Tehn. Estest.} {\bf 7}, 25 (1998).
\item
K. Fujikawa and R.E. Shrock, {\it Phys. Rev. Lett.}
{\bf 45}, 963 (1980).
\item
R.B. Begzhanov and R.S. Sharafiddinov, in {\it Proc. Int. Conf.
on Nuclear Physics}, Moscow, June 16-19, 1998, St. Petersburg, 1998, p. 354.
\item
S. Davidson, B. Campbell and K.D. Bailey, {\it Phys. Rev.}
{\bf  D43}, 2314 (1991); J.A. Morgan and D.B. Farrant,
{\it Phys. Lett.} {\bf B128}, 431 (1983).
\item
Ya. B. Zel'dovich, {\it Zh. Eksp. Teor. Fiz.}
{\bf 33}, 1531 (1957); Ya. B. Zel'dovich and A. M. Perelomov,
{\it ibid.} {\bf 39}, 1115 (1960).
\item
M.J. Musolf and B.R. Holstein,
{\it Phys. Rev.} {\bf D43},1956 (1991).
\item
L.D. Landau, {\it Zh. Eksp. Teor. Fiz.} {\bf 32},405 (1957);
{\it Nucl. Phys.} {\bf 3}, 127 (1957).
\item
B.K. Kerimov and M.Ya. Safin, {\it Izv. Russ. Acad.
Nauk Ser. Fiz.} {\bf 57}, 93 (1993).
\item
S.L. Glashow, {\it Nucl. Phys.} {\bf 22}, 579 (1961);
A. Salam and J.C. Ward, {\it Phys. Lett.} {\bf 13}, 168 (1964);
S. Weinberg, {\it Phys. Rev. Lett.} {\bf 19}, 1264 (1967).
\item
T.W. Donnelly and R.D. Peccei, {\it Phys. Rep.}
{\bf 50}, 3 (1979).
\item
J.Bernstein, M. Ruderman and G. Feinberg, {\it Phys.
Rev.} {\bf 132} 1227 (1963).
\item
R.B. Begzhanov and R.S. Sharafiddinov, in
{\it Proc. Int. Conf. on Nuclear Physics}, Dubna, April 21-24, 1999,
St. Petersburg, 1999, p. 408.
\end{enumerate}
\end{document}